\documentclass[10pt,superscriptaddress,nofootinbib,twocolumn]{revtex4-2}

\usepackage{amsthm,amsmath,amssymb}
\usepackage{mathrsfs}
\usepackage{appendix}
\usepackage{amstext}
\usepackage{graphicx}
\usepackage{url}
\usepackage{float}
\usepackage{bm}
\usepackage{braket}
\usepackage{lineno}
\usepackage[usenames,dvipsnames]{color}
\usepackage[colorlinks=true,citecolor=blue,urlcolor=black]{hyperref}
\usepackage{dcolumn}
\usepackage{natbib}
\usepackage{booktabs}

\begin{document}
\title{Source-independent quantum secret sharing with entangled photon pair networks}

\author{Yi-Ran Xiao}
\author{Zhao-Ying Jia}
\affiliation{National Laboratory of Solid State Microstructures and School of Physics, Collaborative Innovation Center of Advanced Microstructures, Nanjing University, Nanjing 210093, China.}
\affiliation{Department of Physics and Beijing Key Laboratory of Opto-Electronic Functional Materials and Micro-Nano Devices, Key Laboratory of
Quantum State Construction and Manipulation (Ministry of Education), Renmin University of China, Beijing 100872, China.}

\author{Yu-Chen Song}
\affiliation{Big Data Center, Ministry of Emergency Management, Beijing 100013, China}

\author{Yu Bao}
\affiliation{National Laboratory of Solid State Microstructures and School of Physics, Collaborative Innovation Center of Advanced Microstructures, Nanjing University, Nanjing 210093, China.}
\affiliation{Department of Physics and Beijing Key Laboratory of Opto-Electronic Functional Materials and Micro-Nano Devices, Key Laboratory of
Quantum State Construction and Manipulation (Ministry of Education), Renmin University of China, Beijing 100872, China.}

\author{Yao Fu}
\affiliation{Beijing National Laboratory for Condensed Matter Physics and Institute of Physics, Chinese Academy of Sciences, Beijing 100190, China}

\author{Hua-Lei Yin}\email{hlyin@ruc.edu.cn}
\affiliation{Department of Physics and Beijing Key Laboratory of Opto-Electronic Functional Materials and Micro-Nano Devices, Key Laboratory of
Quantum State Construction and Manipulation (Ministry of Education), Renmin University of China, Beijing 100872, China.}

\author{Zeng-Bing Chen}
\affiliation{National Laboratory of Solid State Microstructures and School of Physics, Collaborative Innovation Center of Advanced Microstructures, Nanjing University, Nanjing 210093, China.}
\date{\today}

\begin{abstract}
The large-scale deployment of quantum secret sharing (QSS) in quantum networks is currently challenging due to the requirements for the generation and distribution of multipartite entanglement states. Here we present an efficient source-independent QSS protocol utilizing entangled photon pairs in quantum networks. Through the post-matching method, which means the measurement events in the same basis are matched, the key rate is almost independent of the number of participants. In addition, the unconditional security of our QSS against internal and external eavesdroppers can be proved by introducing an equivalent virtual protocol. Our protocol has great performance and technical advantages in future quantum networks.
\end{abstract}
\maketitle

\section{introduction}
Building a global quantum network would constitute a significant breakthrough in science and technology, as it would extend the benefits of quantum entanglement and quantum state distribution to multiple distant communicating parties worldwide~\cite{wehner2018quantum}. 
With the emergence of quantum technologies, such a network can have numerous applications, including quantum key distribution~\cite{Zhou2023OvercomeRateloss,xie2022breaking,zeng2022mode}, quantum conference key agreement ~\cite{Chen2007GHZQCKA,Li2023BreakingQCKA}, quantum secret sharing (QSS)~\cite{Hillery1999QSS,Cleve1999QSS,Fu2015MDIQSSQCKA},  quantum digital signatures~\cite{Yin2023QDSNetwork}, quantum voting~\cite{vaccaro2007quantum} and quantum secure direct communication~\cite{PhysRevA.65.032302,wen2007secure,zhang2017quantum}.
QSS promises the unconditional security and plays a crucial role in protecting secret information, such as secure operations of multiparty computation~\cite{gottesman1999demonstrating}, creating joint checking accounts containing quantum money~\cite{wiesner1983conjugate} and so on. 

Since the first QSS protocol introduced by Hillery et al.~\cite{Hillery1999QSS} in 1999, which utilizes three-photon Greenberger-Horne-Zeilinger (GHZ) state, this protocol has been generalized into plenty of variations based on entanglement states~\cite{Xiao2004MultipartyQSS}, graph state~\cite{Markham2008GraphQSS,Cai2017MultimodeGraphQSS}, bound entanglement state~\cite{Zhou2018BoundEntangleQSS} and $d$-dimensional entanglement states~\cite{Yu2008MultilevelQSS}. 
In recent decades, significant theoretical~\cite{Singh2005GeneralizedQSS,Tavakoli2015DLevelQSS,Gu2021DPSTFQSS,Li2023AMDIQSS} and experimental~\cite{
Gaertner2007FourPartyQSS,Bell2014ExperimentGraphQSS,Williams2019EntangleQSS,Shen2023PhaseCodingQSS} efforts have been directed toward QSS, making it as one of the most captivating research topics in the quantum cryptography.
Nevertheless, the scheme that distributes multipartite or high-dimensional entanglement resources among multiple participants
renders the aforementioned works impractical due to the limited efficiency and low key rates. 
Additionally, 
adding or removing participants requires changing the dimensionality of the system, which makes complex alterations of the source necessary. 
Therefore, a practical and highly efficient QSS protocol for quantum networks is still missing.

Here, we propose an efficient source-independent QSS protocol utilizing entangled photon pairs, which can achieve data correlation through the post-matching technique~\cite{Lu2021EfficientQDS} and classical XOR operations. 
Based on the fully connected quantum network architecture~\cite{Wengerowsky2018EntangleNetwork,Joshi2020EntangleNetwork,liu202240}, which employs a single entangled photon pair source and wavelength-division multiplexing to distribute bipartite entangled states to multiple users, our scheme can be scaled in multiparty scenarios without changing the entanglement source and the hardware of network users.
The unconditional security of our protocol is proved by introducing an equivalent virtual protocol which utilizes the quantum operations and multiparty entanglement purification~\cite{Dur1999SeparaDistill,maneva2002improved} to obtain the almost pure entangled GHZ states among all QSS participants. 
To address the internal participant attacks~\cite{Karlsson1999QSSParticipantAttack,Qin2007HBBCryptanalysis}, we utilize the method in refs.~\cite{Kogias2017CVQSS,Walk2021Sharing}, where the dealer calculates the correlations with each single player to defend against the internal malicious participants.
The simulation results show that our protocol has a significant advantage of key rate across any number of participants compared with the QSS protocol which directly distributes the GHZ states. 
We believe that our protocol has great potential to serve as a fundamental cryptographic protocol for large-scale quantum networks.

\section{protocol description}
The schematic of our protocol in the fully connected quantum network is shown in Fig.~\ref{Network}.  
In our protocol, the entangled photon pair source generates polarization-entangled photon pairs, i.e., Bell states $\ket{\phi^+} = \left(\ket{HH}+\ket{VV}\right) / \sqrt{2}$, where $\ket{H}$ ($\ket{V}$) represents horizontal (vertical) polarization states.
The diagonal and anti-diagonal polarization states are denoted as $\ket{+}=\left(\ket{H}+\ket{V}\right) / \sqrt{2}$ and $\ket{-}=\left(\ket{H}-\ket{V}\right) / \sqrt{2}$, respectively. 
The network provider distributes the bipartite entangled states between all $N$ network users through the fully connected quantum network architecture~\cite{Wengerowsky2018EntangleNetwork,Joshi2020EntangleNetwork,liu202240}. 
Among the all $N$ network users, $n$ ($n \leq N$) network users who implement the QSS protocol are denoted as QSS participants, which include one dealer $A$ and $n-1$ players $B_i$ ($i = 1, \cdots, n-1$).
All $n$ QSS participants utilize the bipartite entanglement between the dealer $A$ and each player $B_i$ to generate the QSS keys. 
The steps are as follows: 

\begin{enumerate}
\item \textbf{Distribution.} The network provider prepares broadband polarization-entangled photon pairs with a single entangled photon pair source. The spectrum of photon pairs is separated into $N(N-1)$ distinct wavelength channels through a wavelength demultiplexer, and the entangled photon pairs can only be observed in correlated wavelength channels. Then each set of $N-1$ different channels is combined into a single-mode fiber through a wavelength multiplexer and distributed to each network user. Therefore, every network user shares bipartite photon pairs with any other users~\cite{Wengerowsky2018EntangleNetwork,Joshi2020EntangleNetwork,liu202240}.

\item \textbf{Measurement.} All QSS participants perform measurements on their own photons in the diagonal polarization basis, i.e., $X$-basis with probability $p_x$ and the rectilinear polarization basis, i.e., $Z$-basis with probability $p_z=1-p_x$. The measurement results from $\ket{+}$ and $\ket{H}$ are recorded as logic bit 0, and the measurement results from $\ket{-}$ and $\ket{V}$ are are recorded as logic bit 1. 

\item \textbf{Post-matching.} All QSS participants broadcast their basis choices through the classical channels, and the events that the dealer $A$ and any one of the players $B_i$ select the same basis are regarded as matching events. The $X$-basis ($Z$-basis) measurement outcomes of matching events are denoted as $a_{i,j}^X$ ($a_{i,j}^Z$) for the dealer $A$ and $b_{i,j}^X$ ($b_{i,j}^Z$) for each player $B_i$, where $j$ represents the round of matching events.  
Without considering noise, the classical bits $a_{i,j}^X$ ($a_{i,j}^Z$) and $b_{i,j}^X$ ($b_{i,j}^Z$), which are both obtained from the Bell states $\ket{\phi^+}$, should be equal. 

Then the dealer $A$ performs XOR operations on her classical bits. For the $Z$-basis, 
\begin{equation}
	c_{i,j}^Z = a_{1,j}^Z \oplus a_{i,j}^Z,
\end{equation}
and for the $X$-basis, 
\begin{equation}
	\tilde{a}_{1,j}^X = a_{1,j}^X \oplus a_{2,j}^X \oplus \cdots \oplus a_{n-1,j}^X.
\end{equation}
After performing XOR operations, the dealer $A$ broadcasts $c_{i,j}^Z$ through classical channel, and each player $B_i$ also conduct XOR operations $\tilde{b}_{i,j}^Z = c_{i,j}^Z \oplus b_{i,j}^Z$. The value of $\tilde{b}_{i,j}^Z$ is equal to $a_{1,j}^Z$ since $a_{i,j}^Z=b_{i,j}^Z$ and self-inverse of XOR. Owing to the above XOR operations , correlations are established among the classical bits of all participants: 
\begin{align}
	\tilde{a}_{1,j}^X = b_{1,j}^X \oplus b_{2,j}^X \oplus \cdots \oplus b_{n-1,j}^X ,\\
	a_{1,j}^Z = \tilde{b}_{1,j}^Z = \tilde{b}_{2,j}^Z = \cdots = \tilde{b}_{n-1,j}^Z ,
\end{align}
which are equivalent to the GHZ states shared among $n$ participants. 

\item \textbf{Parameter estimation.} After above processing, the classical bits are denoted as $\mathcal{X}=\{\mathcal{X}_A,\mathcal{X}_{B_1},\cdot \cdot \\ \cdots,\mathcal{X}_{B_{n-1}}\}$ and $\mathcal{Z}=\{\mathcal{Z}_A,\mathcal{Z}_{B_1},\cdots,\mathcal{Z}_{B_{n-1}}\}$ for the $X$- and $Z$-bases respectively, where $\mathcal{X}_A$ ($\mathcal{Z}_A$) represents the classical bit string $\{\tilde{a}_{1,j}^X\}$ ($\{a_{1,j}^Z\}$) of the dealer $A$ and $\mathcal{X}_{B_i}$ ($\mathcal{Z}_{B_i}$) represents the classical bit string $\{b_{i,j}^X\}$ ($\{\tilde{b}_{i,j}^Z\}$) of each player $B_i$. To conduct parameter estimation, the dealer $A$ would designate the classical bits in $\mathcal{Z}$ for disclosure. Above steps are repeated until at least $m$ rounds are kept for key generation and $k$ rounds are kept for parameter estimation. The dealer $A$ and each single player $B_i$ then utilize the classical bits in $\mathcal{Z}$ to calculate the correlation between them. Based on the correlations, the dealer $A$ decides whether the protocol is aborted or not.

\begin{figure}[t]
\centering
\includegraphics[width=\linewidth]{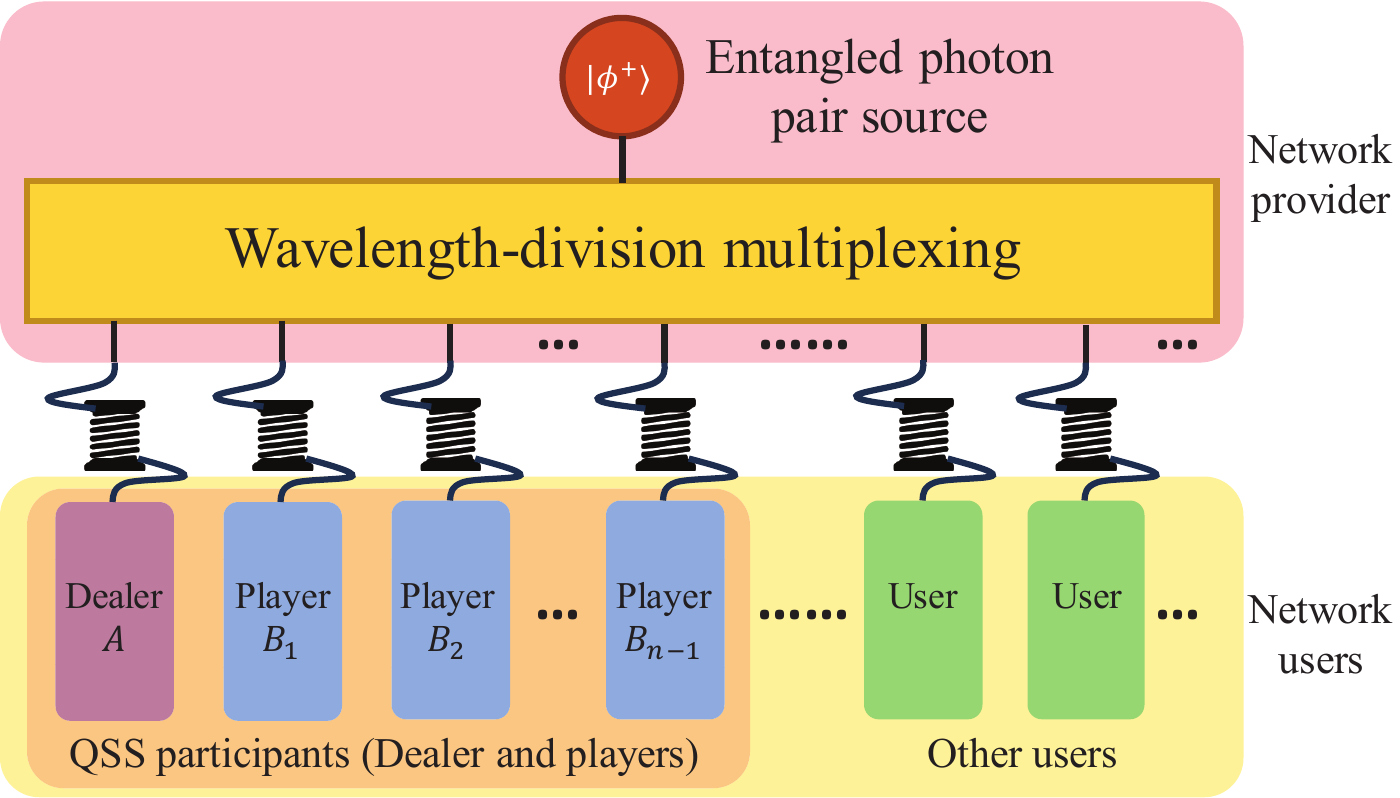}
\caption{
The schematic of our QSS protocol in quantum network.
The network provider (red box) uses a single entangled photon pair source to distribute bipartite entangled states between all network users (yellow box) through wavelength-division multiplexing~\cite{Wengerowsky2018EntangleNetwork,Joshi2020EntangleNetwork,liu202240}. 
Some of the network users implement the QSS protocol, thus they are denoted as QSS participants, including the dealer and the players. 
The QSS participants (orange box) utilize the bipartite entanglement to generate secure keys with the post-matching technique~\cite{Lu2021EfficientQDS} and classical XOR operations.
} 
\label{Network}
\end{figure}

\item \textbf{Post processing.} If the protocol is not aborted, QSS participants will obtain the correlated raw keys which are equivalent to the noisy GHZ states. All participants proceeds with the error correction, which will leak at most $\mathrm{leak}_{\mathrm{EC}}$ bits of information. To test the correctness, all QSS participants compute and compare a hash of length $\log_2(1/\epsilon_{\mathrm{c}})$ bits by employing a random universal$_2$ hash function on the raw keys. If passing the correctness test, all participants perform privacy amplification using universal$_2$ hashing, which will extract $l$ bits final keys. 
\end{enumerate}

To prove the security of our QSS protocol, we introduce an equivalent virtual protocol, which can be transformed into the actual protocol above. 
In the virtual protocol, the Bell states kept in quantum memories of the dealer $A$ and each player $B_i$ are correspond to the measurement results of entangled photon pairs stored in classical memories. 
The quantum operations conducted on the qubits in quantum memories to generate noisy GHZ states are equivalent to the classical XOR operations performed on classical bits. 
Parameter estimations in both the actual and virtual protocols are conducted only between the dealer $A$ and each single player $B_i$ to exclude the effects of potential dishonest participants~\cite{Kogias2017CVQSS,Walk2021Sharing}.
Multiparty entanglement purification~\cite{Dur1999SeparaDistill,maneva2002improved} is performed on noisy GHZ states to distill nearly pure GHZ states, which can be converted into classical post-processing (classical error correction and privacy amplification)~\cite{Bennett1996EPPandQECC,Shor2000BB84SecurityProof}. 
Due to the above-mentioned equivalence between the two protocols, all participants in the actual protocol can obtain unconditionally secure QSS keys against both the internal and external eavesdroppers, including the internal participant attacks~\cite{Karlsson1999QSSParticipantAttack,Qin2007HBBCryptanalysis}. 
The detailed virtual protocol and security analysis are shown in the Appendix~\ref{Security Analysis}.

\begin{figure}[t]
\centering
\includegraphics[width=\linewidth]{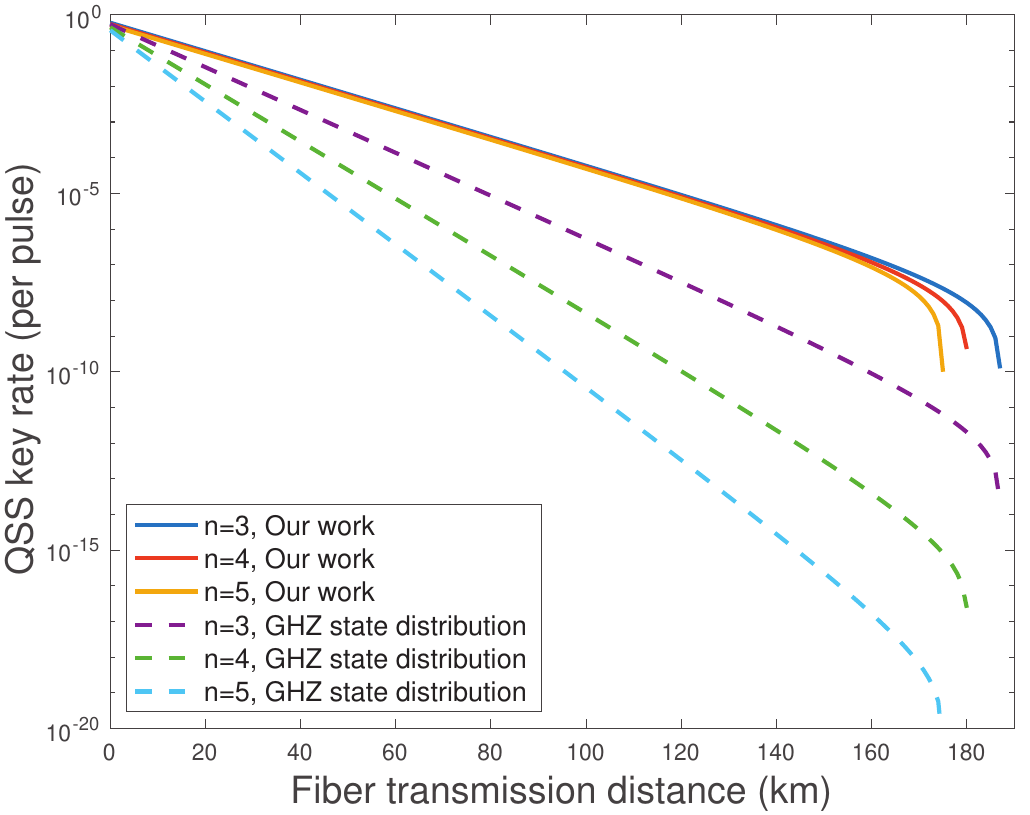}
\caption{
Comparison of our $n$-party protocol and the QSS protocol with GHZ state distribution. 
The key rates of our work (solid lines) and the QSS protocol with GHZ state distribution (dotted lines) are plotted with the different numbers of participants ($n$ = 3, 4, 5). 
The fiber transmission distance denotes the distance between the network provider and one participant.
}
\label{Asymptotic_rate}
\end{figure}

\begin{figure}[t]
\centering
\includegraphics[width=\linewidth]{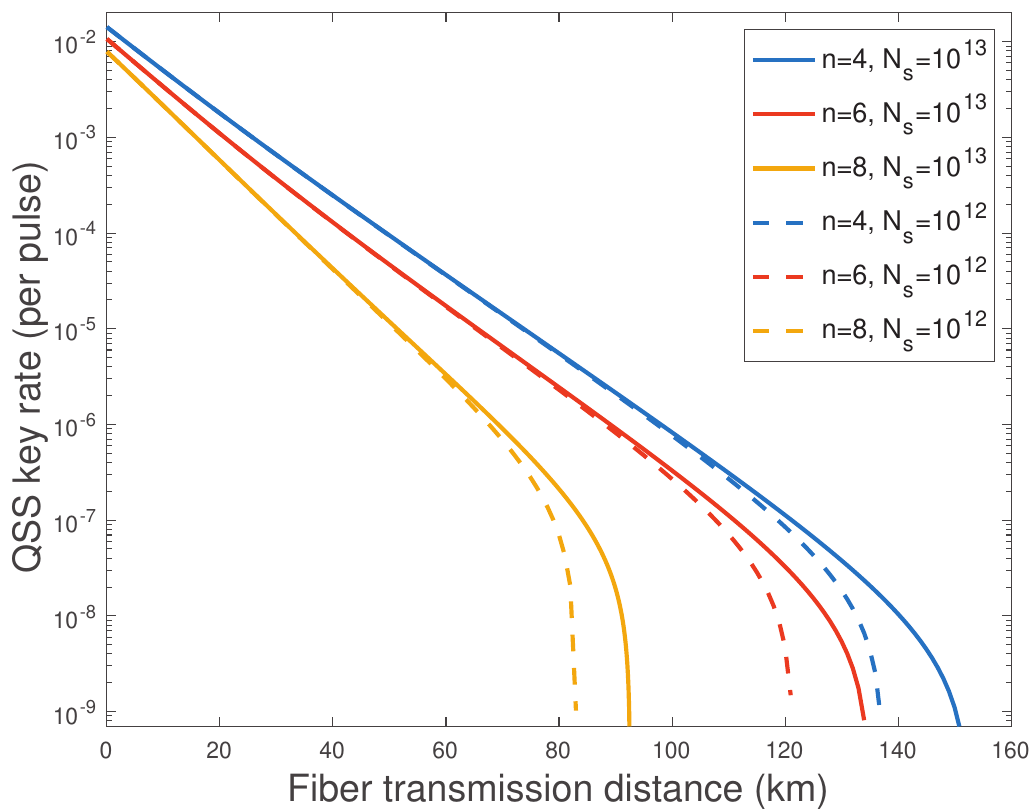}
\caption{
Performance of our $n$-party QSS protocol in finite key regime. 
The finite key rates of our protocol are plotted under the different numbers of participants ($n$ = 4, 6, 8) and the different numbers of signal distributed to each player ($N_s$ =  $10^{12}$, $10^{13}$). 
The fiber transmission distance denotes the distance between the network provider and one participant.
}
\label{Finite_rate}
\end{figure}

\section{numerical simulation}
Based on the actual protocol description above, all QSS participants generate secret keys utilizing the measurement results of nearly perfect GHZ states in the $X$-basis. 
Therefore, we can give the following key rate in the case of asymptotic limit~\cite{Fu2015MDIQSSQCKA,Li2023AMDIQSS}:  
\begin{equation}
	R_{\mathrm{QSS}} = Q^{X} \left[1 - \mathop{\max}\limits_{i} h\left(E^Z_{AB_i}\right) - f_e h\left(E^X\right)\right],
\label{asymptotic rate}
\end{equation}
where $Q^{X}$ is the gain of the $X$-basis, i.e., the probability that both the dealer $A$ and each player $B_i$ receive and measure photons in the $X$-basis. 
$E^X$ is the bit error rate in the $X$-basis. 
$E^Z_{AB_i}$ is the marginal error rate between the dealer $A$ and any single player $B_i$ in correlation test. 
$h(x)=-x\log_{2}(x)-(1-x)\log_{2}(1-x)$ is the binary Shannon entropy function and $f_e$ is the error correction inefficiency. 
The detailed analysis for the gain and error rate is given in the Appendix~\ref{Gain and Error rate}. 

To show the performance of our protocol, we consider a QSS protocol based on GHZ states distribution, which also utilizes measurement results in the $X$-basis for key generation and the $Z$-basis for parameter estimation, and compare the key rate with our work in asymptotic limit. 
For simplicity, we assume that the GHZ source of the network provider is a perfect source which distributes one pure GHZ state at a time, and the transmission distances from the network provider to all participants are equal.
For all participants, the detection efficiency and dark count rate of photon detectors are set as $\eta_d = 90\%$ and $p_d = 10^{-5}$, and the misalignment error rate is $e_d = 0.01$. 
The attenuation coefficient of channels and the error correction inefficiency are fixed as $\alpha = 0.2$ and $f_e = 1.22$, respectively. 
The asymptotic key rate is the same as the Eq.~(\ref{asymptotic rate}), except for the gain (see the Appendix~\ref{Gain and Error rate}).

For comparison, we also assume that the entangled photon pair source of the network provider in our protocol is a perfect source, and utilize the same simulation parameters.
Figure~\ref{Asymptotic_rate} shows that the key rate of our work is almost independent of the number of participants and has a significant advantage compared with the QSS protocol based on GHZ states distribution. 
Let $\eta$ ($\eta < 1$) be the probability that one participant receives a photon, owing to the Bell state distribution and the post-matching technique~\cite{Lu2021EfficientQDS}, the gain is increased from $\mathcal{O}\left(\eta^{n} \right)$ to $\mathcal{O}\left(\eta^2 \right)$ for $n$ QSS participants (see the Appendix~\ref{Gain and Error rate}), thus the high efficiency and robustness of our protocol can be achieved.

In addition, we investigate the key rate in a realistic case, where the finite resource and the realistic entangled photon source are taken into consideration. 
In finite key regime, based on the method of composable security in refs.~\cite{Tomamichel2012QKDFinitekeyAnalysis,Walk2021Sharing,Li2023AMDIQSS} and the statistical fluctuation analysis of random sampling without replacement in ref.~\cite{Yin2020TightSecurityBounds}, we can give the following length $l$ of secure key:
\begin{equation}
\begin{split}
	l = & m \left[q - \mathop{\max}\limits_{i} h\left(E^Z_{AB_i} + \lambda\left(E^Z_{AB_i},\epsilon\right)\right)\right] \\
        & - \mathrm{leak}_{\mathrm{EC}} - \log _{2} \frac{1}{4 \epsilon_{\mathrm{c}} \left( \epsilon^\prime \right)^2} ,
\end{split}
\label{finite rate}
\end{equation}
where 
\begin{footnotesize}
\begin{equation*}
\begin{split}
    \lambda\left(E^Z_{AB_i},\epsilon\right) = \frac{\frac{\left(1-2E^Z_{AB_i}\right)AG}{m+k_i} + \sqrt{\frac{A^2G^2}{\left(m+k_i\right)^2} + 4E^Z_{AB_i}\left(1-E^Z_{AB_i}\right)G}}{2 + 2\frac{A^2G}{\left(m+k_i\right)^2}} , 
\end{split}
\end{equation*}
\end{footnotesize}
$k_i$ ($k_i \leq k$) is the number of correlation test rounds between the dealer $A$ and the $i$th player $B_i$ and $E^Z_{AB_i}$ is the marginal error rate observed in correlation test. 
$A = \max\{m, k_i\}$ and $G = \frac{m+k_i}{mk_i} \ln \frac{m+k_i}{2\pi mk_i E^Z_{AB_i}\left(1-E^Z_{AB_i}\right) \epsilon^2}$. 
$q$ quantifies the complementarity of the two measurement bases, i.e., the $X$- and $Z$-bases.  
We fix $\epsilon_{\mathrm{c}}=3\epsilon=3\epsilon^\prime=10^{-10}$, $p_x=0.9$ and the average number of photon pairs generated per pulse $\mu=0.04$, and assume that the information leakage $\mathrm{leak}_{\mathrm{EC}} = m f_e h\left(E^X\right)$, where $E^X$ is the total error rate in the $X$-basis.
The other simulation parameters are the same as that in asymptotic limit.
According to the results in Fig.~\ref{Finite_rate}, the transmission distance of our QSS protocol is over 130km, 110km and 80km for $n=4,6,8$, respectively, making it suitable for the metropolitan quantum network.
The simulation also shows that the finite key rate of our work decreases as the number of participants $n$ increases, owing to the fact that more participants lead to a higher error rate. 

\section{discussion}
In conclusion, we present an efficient multiparty QSS protocol with entangled photon pair source. 
Note that our QSS protocol is naturally source-independent since it is entanglement-based~\cite{Koashi2003SourceIndependent, Yin2020EntangleQKD1120}. 
Through the post-matching technique~\cite{Lu2021EfficientQDS} and classical XOR operations, the GHZ-state correlation can be established among the measurement results of entangled photon pairs in our protocol. 
Thus, under the architecture of the fully connected quantum networks~\cite{Wengerowsky2018EntangleNetwork,Joshi2020EntangleNetwork,liu202240}, our scheme can be easily extended without any hardware modifications for the network provider and all network users, which is highly versatile and suitable for the metropolitan network. 
By introducing an equivalent virtual protocol, we prove the unconditional security of our QSS protocol against the internal and external eavesdroppers. 
The simulation results show that the key rate of our protocol is nearly independent of the number of participants and has a significant advantage compared to the QSS protocol that directly distributes GHZ states.
The robustness and high efficiency are due to the gain improving from $\mathcal{O}\left(\eta^{n} \right)$ to $\mathcal{O}\left(\eta^2 \right)$ and the use of marginal error rate in $Z$-basis to estimate the phase error rate (see the Appendix~\ref{Gain and Error rate}).
Moreover, our protocol can be readily implemented with current quantum key distribution devices and entanglement sources. 
Therefore, combined with the high efficiency quantum cryptographic protocol, for example, the efficient quantum digital signatures protocol~\cite{Li2023OTUHQDS}, our protocol enable numerous potential applications to be realized, such as quantum Byzantine agreement~\cite{weng2023beating,jing2024experimental} and quantum e-commerce~\cite{Cao2024Qecommerce} in quantum networks.
We anticipate that our work will serve the infrastructure of large-scale deployment for quantum networks.

\section*{Acknowledgments}
We gratefully acknowledge the supports from the National Natural Science Foundation of China (No. 12274223); Program for Innovative Talents and Entrepreneurs in Jiangsu (No. JSSCRC2021484).

\appendix
\section{Post-matching method}

To describe the post-matching technique~\cite{Lu2021EfficientQDS} in our source-independent quantum secret sharing (QSS) protocol in more detail, we present the specific steps below.

\begin{enumerate}
\item After measuring the photons in $X$-basis and $Z$-basis with corresponding probability, all QSS participants broadcast their basis choices. The measurement events that the dealer $A$ and any one of the players $B_i$ select the same basis are regarded as matching events $M_{i,k}^S$, where $S \in \{X, Z\}$ represents the measurement basis, $i$ represents the event between the dealer $A$ and the players $B_i$, $k$ represents the round of measurement events.

\item The matching events $M_{i,k}^S$ in the same measurement basis are matched in sequence, and the measurement results in different bases are categorized into different groups. Therefore, the raw data $\{a_{i,j}^S\}$ can be obtained for the dealer $A$ and $\{b_{i,j}^S\}$ are obtained for the player $B_i$, where $j$ represents the round of matching events. 
\end{enumerate}

\begin{figure}[htbp]
\centering
\includegraphics[width=\linewidth]{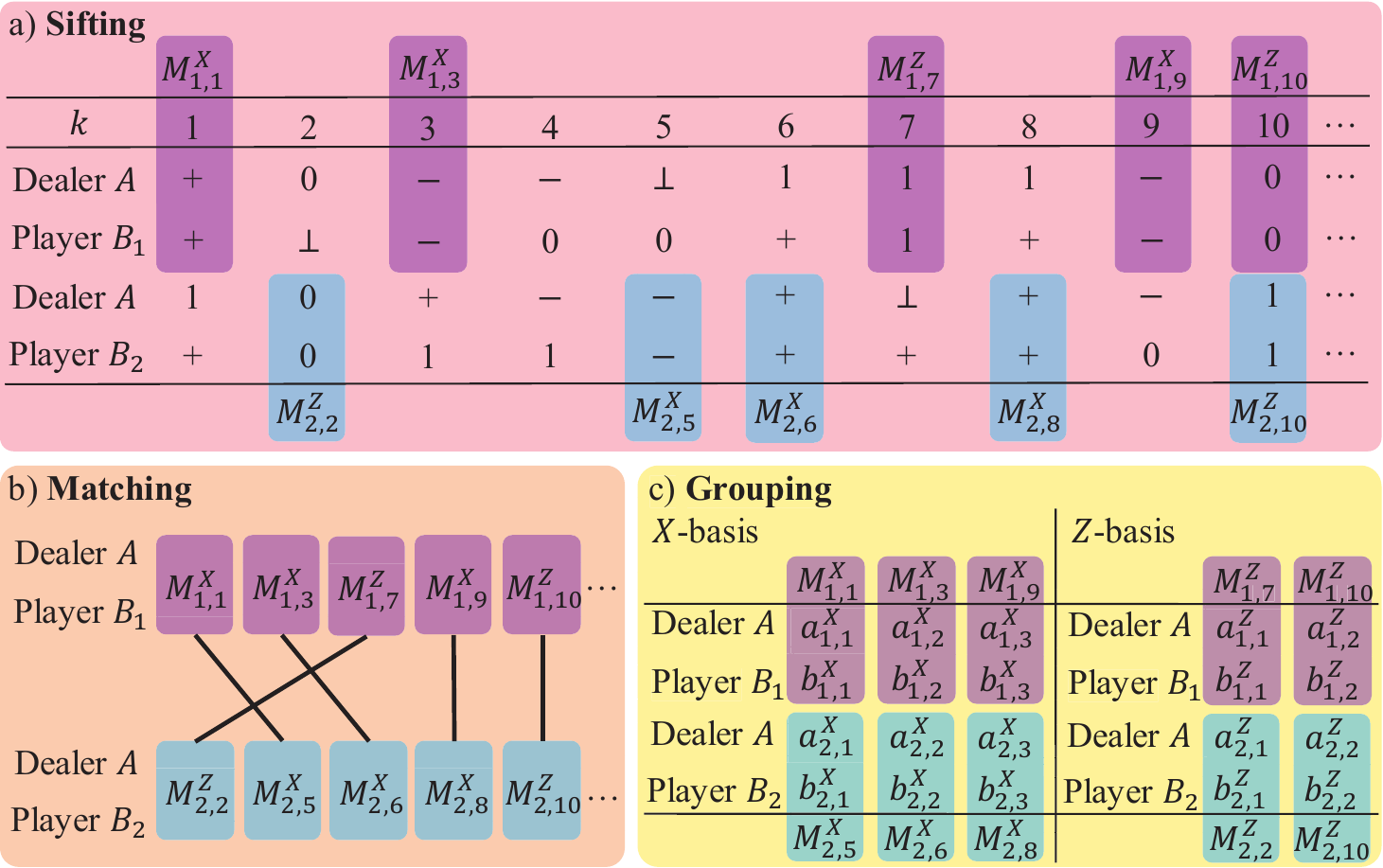}
\caption{
The schematic for the process of the post-matching method in the three-party case.
For simplicity, we assumed the measurement results of the first ten events.
$\{0, 1\}$ represents the measurement results in $Z$-basis. 
$\{+, -\}$ represents the measurement results in $X$-basis, where $+$ corresponds to the logic bit $0$ and $-$ corresponds to the logic bit $1$. 
$\bot$ represents no measurement result. }
\label{Post_matching}
\end{figure}

Here, we provide a post-matching example in three-party case, which includes the dealer $A$ and the players $B_1$ and $B_2$, to clarify the process above. As shown in Fig.~\ref{Post_matching}, the matching events between the dealer $A$ and the player $B_1$ are $\{M_{1,1}^X, M_{1,3}^X, M_{1,7}^Z, M_{1,9}^X, M_{1,10}^Z, \cdots\}$, and the matching events between the dealer $A$ and the player $B_2$ are $\{M_{2,2}^Z, M_{2,5}^X, M_{2,6}^X, M_{2,8}^X, M_{2,10}^Z, \cdots\}$. Then the events $M_{1,1}^X$ and $M_{2,5}^X$ are matched, and the raw data $a_{1,1}^X, a_{2,1}^X$ are obtained for dealer $A$ and $b_{1,1}^X$ ($b_{2,1}^X$) is for the player $B_1$ ($B_2$). Similarly, the other events are matched in sequence and the raw data can be obtained for all participants. 

\section{Security analysis}
\label{Security Analysis}

\subsection{Virtual protocol}

\begin{figure}[htbp]
\centering
\includegraphics[width=\linewidth]{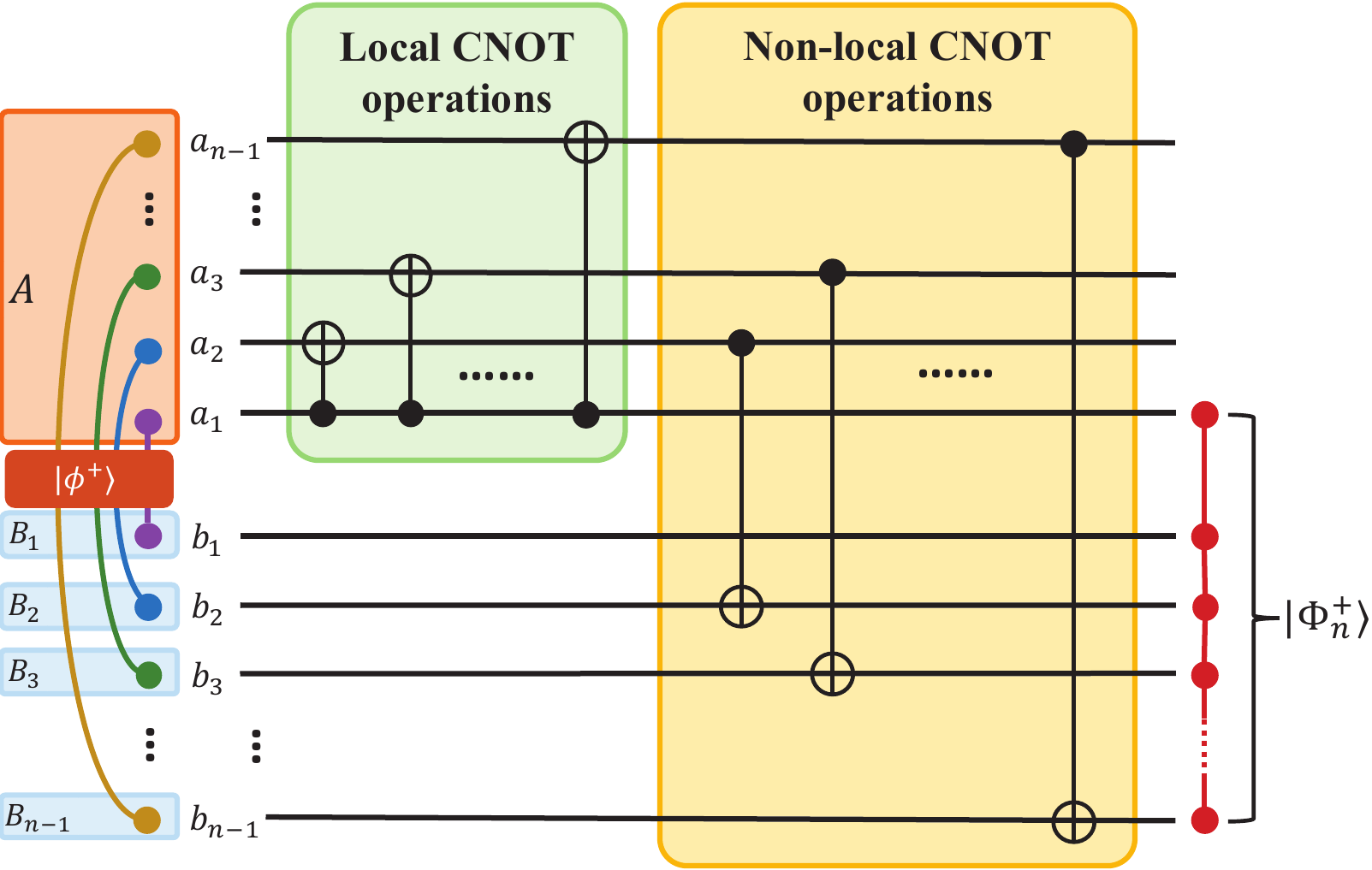}
\caption{
The schematic for the generation of $n$-party GHZ state through quantum operations, which are divided into two parts: local CNOT operations of the dealer $A$ and non-local CNOT operations between the dealer $A$ and each player $B_i$. 
Different Bell states $\ket{\phi^+}$ are represented in different colors.
The virtual qubits $\{a_1,b_1,b_2,\cdots,b_{n-1}\}$ are transformed into $n$-party GHZ states $\ket{\Phi_n^+}$.
}
\label{Virtual_Protocol}
\end{figure}

To prove the security of our QSS protocol, we introduce a virtual protocol, in which the QSS participants establish entanglement of GHZ state among multiple Bell states by controlled-NOT (CNOT) operations, which are shown in Fig.~\ref{Virtual_Protocol}, and obtain almost pure Greenberger-Horne-Zeilinger (GHZ) state through the entanglement purification. The details of the virtual protocol are as follows:

\begin{enumerate}
\item The entanglement source of the network provider generates $n-1$ Bell states $\ket{\phi^+} = \left(\ket{00}+\ket{11}\right) / \sqrt{2}$, and each state contains one virtual qubit $a_i$ kept in quantum memory of the dealer $A$ and the other virtual qubit $b_i$ is sent to each player $B_i$ respectively. After receiving the virtual qubits, all players store their own qubits in quantum memory for the following operations. 

\item The dealer $A$ first uses virtual qubit $a_1$ as the control bit to perform multiple CNOT operations on her remaining virtual qubits $a_i$ ($i=2,3,\cdots,n-1$). After the local CNOT operations, the dealer $A$ and each player $B_i$ perform a non-local CNOT operation with virtual qubit $a_i$ as control qubit and virtual qubit $b_i$ as target qubit, except for the player $B_1$. After the processing above, the virtual qubits of all parties $\{a_1,b_1,b_2,\cdots,b_{n-1}\}$ are transformed into an $n$-party GHZ state $\ket{\Phi^+_{n}} = \frac{1}{\sqrt{2}} \left(\ket{0}^{\otimes n}+\ket{1}^{\otimes n}\right)$. The detailed evolution of quantum operations is shown in Appendix~\ref{CNOT and XOR}.

\item After repeating two steps above for sufficient times, all participants share a set of noisy $n$-party GHZ states. Then a random subset of GHZ states is selected by the dealer $A$ and measured in the $Z$-basis for parameter estimation. The dealer $A$ and each single player $B_i$ utilize the measurement outcomes above to calculate the correlation between them, and the protocol aborts if the correlations are below a certain level. 

\item If the correlation test passes, the entanglement purification is performed among all noisy GHZ states to distillate approximately pure GHZ states. The dealer $A$ and all $n-1$ players $B_i$ ($i=1,2,\cdots,n-1$) perform measurement in the $X$-basis on the GHZ states and utilize the outcomes to generate the final key for QSS.

\end{enumerate}
    
The quantum operations above performed on the virtual qubits are corresponding to the XOR operations performed on the classical bits in the actual protocol, where the local CNOT operations are equivalent to the XOR operations conducted by the dealer $A$ on her measurement outcomes in the $X$- and $Z$-bases, and the non-local CNOT operations correspond to the broadcast of the dealer $A$ and XOR operations conducted by all players. The details of correspondence are shown in Appendix~\ref{CNOT and XOR}. 
Entanglement purification in the virtual protocol can be converted into the quantum error correction~\cite{Bennett1996EPPandQECC}, and the security of quantum communication protocols can be proved utilizing Calderbank-Shor-Steane (CSS) code~\cite{Shor2000BB84SecurityProof}. Then the property of CSS code allows for the transformation of quantum error correction into classical post-processing, which leads to the bit error correction (phase error correction) can be regarded as the classical error correction (privacy amplification) in the actual protocol. 
Parameter estimations in both the virtual protocol and actual protocol are conducted only between the dealer $A$ and each single player $B_i$ to exclude the effects of potential dishonest parties. Following the Refs.~\cite{Kogias2017CVQSS,Walk2021Sharing}, parameter estimation performed between the dealer and each complementary party (single player in our protocol) corresponding to unauthorized subset can against internal participant attacks~\cite{Karlsson1999QSSParticipantAttack,Qin2007HBBCryptanalysis}, which has been proven as a loophole in the security of original QSS protocol~\cite{Hillery1999QSS}.

Therefore, the virtual protocol is equivalent to our actual protocol, which indicates that all participants in the actual protocol can share the classical bits with the correlation of almost pure GHZ state. 
Owing to the monogamy of entanglement~\cite{Terhal2004EntangleMonogamous}, the information leaked to eavesdropper is negligible and all parties can obtain an information-theoretically secure key by sharing almost perfect GHZ states in the virtual protocol.
Therefore, our actual protocol can also avoid the risk of information leakage and obtain a secure key for QSS. 

\subsection{Quantum operations and classical XOR operations}
\label{CNOT and XOR}

The schematic diagram of quantum operations that generating $n$-party GHZ state is shown in Fig.~\ref{Virtual_Protocol}. Here we represent the Bell state distributed between the dealer $A$ and the player $B_i$ in $\sigma_z$ representation: 
\begin{equation}
	\ket{\phi^+}_{a_i b_i} = \frac{1}{\sqrt{2}}\left(\ket{00}_{a_i b_i} + \ket{11}_{a_i b_i}\right)=\frac{1}{\sqrt{2}}\sum_{z_i = 0}^1 \ket{z_i z_i}_{a_i b_i},
\label{Bell state}
\end{equation}
where $a_i$ and $b_i$ denote qubits belong to the dealer $A$ and the player $B_i$ respectively. After the distribution of Bell states, the quantum state among all $n$ participants can be denoted as 
\begin{widetext}
\begin{equation}
\begin{split}
	\ket{\Psi} &= \ket{\phi^+}_{a_1 b_1} \ket{\phi^+}_{a_2 b_2} \cdots \ket{\phi^+}_{a_{n-1} b_{n-1}}\\
	&= \frac{1}{2^{(n-1)/2}} \left(\sum_{z_1=0}^{1} \ket{z_1 z_1}_{a_1 b_1} \right) \left(\sum_{z_2=0}^{1} \ket{z_2 z_2}_{a_2 b_2} \right) \cdots \left(\sum_{z_{n-1}=0}^{1} \ket{z_{n-1} z_{n-1}}_{a_{n-1} b_{n-1}} \right)\\
	&= \frac{1}{2^{(n-1)/2}} \sum_{\substack{\{z_i\}\\z_i \in[0,1]}} \ket{z_1 z_1 z_2 z_2 \cdots z_{n-1} z_{n-1}}_{a_1 b_1 a_2 b_2 \cdots a_{n-1} b_{n-1}},
\end{split}
\label{n-party Bell state}
\end{equation}
where the summation symbol in the third line of Eq.~(\ref{n-party Bell state}) represents the sum over all possible values of the sequence $\{z_i|i=1,2,\cdots,n-1\}$. According to the action of CNOT operation $\ket{c}\ket{t}\rightarrow\ket{c}\ket{t \oplus c}$, the dealer $A$ first performs local CNOT operations on her virtual qubits 
\begin{equation}
	\ket{z_1}_{a_1}\ket{z_i}_{a_i} \rightarrow \ket{z_1}_{a_1}\ket{z_i \oplus z_1}_{a_i},
\end{equation}
and then conducts non-local CNOT operations with each player $B_i$, except for $B_1$ 
\begin{equation}
	\ket{z_i \oplus z_1}_{a_i}\ket{z_i}_{b_i} \rightarrow \ket{z_i \oplus z_1}_{a_i}\ket{z_i \oplus z_i \oplus z_1}_{b_i} = \ket{z_i \oplus z_1}_{a_i}\ket{z_1}_{b_i}
\end{equation}
Therefore, the quantum state among $n$ parties goes to 
\begin{equation}
\begin{split}
	\ket{\Psi} \xrightarrow[\mathrm{CNOT}]{\mathrm{Local}}& \frac{1}{2^{(n-1)/2}} \sum_{\substack{\{z_i\}\\z_i \in[0,1]}} \ket{z_1 z_1 (z_2 \oplus z_1) z_2 \cdots (z_{n-1} \oplus z_1) z_{n-1}}_{a_1 b_1 a_2 b_2 \cdots a_{n-1} b_{n-1}}\\
	\xrightarrow[\mathrm{CNOT}]{\mathrm{Non-local}}& 
	\frac{1}{2^{(n-1)/2}} \sum_{\substack{\{z_i\}\\z_i \in[0,1]}} \ket{z_1 z_1 (z_2 \oplus z_1) z_1 \cdots (z_{n-1} \oplus z_1) z_1}_{a_1 b_1 a_2 b_2 \cdots a_{n-1} b_{n-1}}\\
	=& \frac{1}{2^{(n-1)/2}} \sum_{\substack{\{z_i\}\\z_i \in[0,1]}} \ket{z_1}^{\otimes n}_{a_1 b_1 b_2 \cdots b_{n-1}} \ket{(z_2 \oplus z_1)\cdots (z_{n-1} \oplus z_1)}_{a_2 \cdots a_{n-1}}\\
	=& \frac{1}{2^{(n-1)/2}} \sum_{\substack{\{z_i|i \neq 1\}\\z_i \in[0,1]}} \left(\ket{0}^{\otimes n}_{a_1 b_1 b_2 \cdots b_{n-1}} \ket{z_2 \cdots z_{n-1}}_{a_2 \cdots a_{n-1}} + \ket{1}^{\otimes n}_{a_1 b_1 b_2 \cdots b_{n-1}} \ket{\bar{z}_2 \cdots \bar{z}_{n-1}}_{a_2 \cdots a_{n-1}} \right),
\end{split}
\label{CNOT operations}
\end{equation}
where $\bar{z}_i$ represents the negation of $z_i$. The value of $\bar{z}_i$ at $z_i=0$ ($z_i=1$) is the same as the value of $z_i$ at $z_i=1$ ($z_i=0$), therefore the summation of all possible values of sequence $\{\bar{z}_i|i\neq1\}$ is equal to that of sequence $\{z_i|i\neq1\}$, which gives the quantum state as
\begin{equation}
\begin{split}
	\ket{\Psi} \xrightarrow[\mathrm{operations}]{\mathrm{Quantum}}& \frac{1}{2^{(n-1)/2}} \sum_{\substack{\{z_i|i \neq 1\}\\z_i \in[0,1]}} \ket{z_2 \cdots z_{n-1}}_{a_2 \cdots a_{n-1}} \left(\ket{0}^{\otimes n} + \ket{1}^{\otimes n}\right)_{a_1 b_1 b_2 \cdots b_{n-1}}\\
	=& \frac{1}{2^{(n-1)/2}} \left(\ket{0}+\ket{1}\right)_{a_2} \cdots \left(\ket{0}+\ket{1}\right)_{a_{n-1}} \left(\ket{0}^{\otimes n} + \ket{1}^{\otimes n}\right)_{a_1 b_1 b_2 \cdots b_{n-1}}.
\end{split}
\label{before GHZ state}
\end{equation}
\end{widetext}
Then an $n$-party GHZ state is established among the subsystem $\{a_1 b_1 b_2 \cdots b_{n-1}\}$ of all participants
\begin{equation}
	\ket{\Phi^+_n} = \frac{1}{\sqrt{2}} \left(\ket{0}^{\otimes n} + \ket{1}^{\otimes n}\right)_{a_1 b_1 b_2 \cdots b_{n-1}}.
\label{GHZ state}
\end{equation}

The process of GHZ states generation above should be achieved by quantum memories and multiple quantum gates with high fidelity, which are far from practical application. 
Therefore, in the actual protocol, we utilize the classical XOR operations to achieve the correlation of GHZ state among the measurement results in the $X$- and $Z$-bases.
The corresponding relationship between the quantum operations and classical operations is shown below. 

We first recall the action of CNOT operation in the $Z$- and $X$-bases. For the $Z$-basis, the action of CNOT operation is flipping target qubit according to control qubit, i.e., $\ket{c}\ket{t}\rightarrow\ket{c}\ket{t \oplus c}$. 
However, for the $X$-basis, the CNOT behavior is changed: target qubit is not flipped, while control qubit is flipped if the state of target qubit is $\ket{-} = (\ket{0} - \ket{1}) / \sqrt{2}$ and remains unchanged if the state of target qubit is $\ket{+} = (\ket{0} + \ket{1}) / \sqrt{2}$, i.e., $\ket{c}\ket{t}\rightarrow\ket{c \oplus t}\ket{t}$. 

Therefore, the local CNOT operations performed by the dealer $A$ corresponds to the following classical XOR operations.
Note that the following operations are conducted on the same round of matching events, thus we neglect the index $j$, which represents the round of matching events. 
For the $Z$-basis, the measurement results $a_{i}^Z$ ($i=2,3,\cdots,n-1$) are replaced with $c_{i}^Z = a_{1}^Z \oplus a_{i}^Z$ but $a_{1}^Z$ remains unchanged. 
For the $X$-basis, $a_{1}^X$ is changed into $\tilde{a}_{1}^X = a_{1}^X \oplus a_{2}^X \oplus \cdots \oplus a_{n-1}^X$ and the other bits are not changed. 
The non-local CNOT operations between the dealer $A$ and each player $B_i$ can also be transformed into the classical process below. 
For the $Z$-basis, the dealer $A$ broadcasts $c_{i}^Z$ and each player $B_i$ decides whether the measurement results $b_{i}^Z$ is flipped or not according to $c_{i}^Z$, which means $b_{i}^Z$ is changed into $\tilde{b}_{i}^Z = c_{i}^Z \oplus b_{i}^Z$. 
Note that $c_{1}^Z = a_{1}^Z \oplus a_{1}^Z = 0$, i.e., $\tilde{b}_{1}^Z = b_{1}^Z$, which corresponds that the virtual qubit $b_1$ is unchanged under the non-local CNOT operations.
For the $X$-basis, however, the target qubits remains unchanged, which indicates that no classical operations are performed on $b_{i}^X$. 
Here, we should note that the equivalence between the CNOT operations and classical XOR operations shown above is in the noiseless case. 
However, multiparty entanglement purification~\cite{Dur1999SeparaDistill,maneva2002improved} can distill nearly perfect GHZ states from noisy GHZ states, which means the errors during the distribution and CNOT operations can be nearly corrected, and the entanglement purification can be converted into thetran classical error correction and privacy amplification due to the property of CSS code.
Therefore, the virtual protocol is equivalent to the actual protocol in the noisy case.

\begin{figure*}[htbp]
\centering
\includegraphics[width=.8\linewidth]{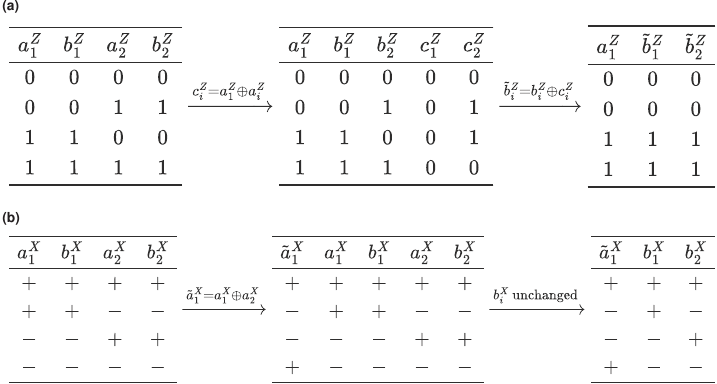}
\caption{The process of classical operations performed on the measurement results in the $Z$- and $X$-bases. (a) For the $Z$-basis, $a_1^Z$ ($a_2^Z$) and $b_1^Z$ ($b_2^Z$) represent the $Z$-basis measurement results of the Bell states distributed between the dealer $A$ and the player $B_1$ ($B_2$). 
(b) For the $X$-basis, $a_1^X$ ($a_2^X$) and $b_1^X$ ($b_2^X$) represent the $X$-basis measurement results of the Bell states distributed between the dealer $A$ and the player $B_1$ ($B_2$). $+$ and $-$ represent the logic bits $0$ and $1$ in the $X$-basis respectively.}
\label{XOR}
\end{figure*}

To clarify the process of XOR operations above, we utilize the case of three parties $\{A, B_1, B_2\}$ as an example and list all possible measurement results in the $Z$- and $X$-bases without noise. 
For the $Z$-basis, the process of XOR operations is shown in Fig.~\ref{XOR} (a). 
In the third table, the correlation of classical bits $\{a_1^Z, \tilde{b}_1^Z, \tilde{b}_2^Z\}$ is $a_1^Z=\tilde{b}_1^Z=\tilde{b}_2^Z$, which is the same as GHZ-state correlation in the $Z$-basis. 
For the $X$-basis, the process is listed in Fig.~\ref{XOR} (b), where the final correlation of classical bits $\{\tilde{a}_1^X, b_1^X, b_2^X\}$ is $\tilde{a}_1^X=b_1^X \oplus b_2^X$, which is the same as GHZ state correlation in the $X$-basis. 
Clearly, the classical XOR operations are corresponding to the quantum CNOT operations and can implement GHZ-state correlation among the classical bits of all QSS participants.

\section{Analysis of the gain and the error rate}
\label{Gain and Error rate}

\subsection{Theoretical analysis}

According to ref.~\cite{Ma2007Entanglement}, the gain and error rate of bipartite entanglement distribution between the dealer $A$ and each player $B_i$ are given as 
\begin{widetext}
\begin{equation}
	Q_\mu = 1 - \frac{1-Y_{0A}}{\left(1+\eta_{A}\frac{\mu}{2}\right)^2} - \frac{1-Y_{0B_i}}{\left(1+\eta_{B_i}\frac{\mu}{2}\right)^2} + \frac{(1-Y_{0A}) (1-Y_{0B_i})}{\left(1+\eta_{A}\frac{\mu}{2}+\eta_{B_i}\frac{\mu}{2}-\eta_{A}\eta_{B_i}\frac{\mu}{2}\right)^2} ,
\end{equation}
\begin{equation}
	E_\mu^{X(Z)} Q_\mu = e_0 Q_\mu - \frac{2 \left(e_0 - e_d^{X(Z)}\right) \eta_{A} \eta_{B_i} \frac{\mu}{2} \left(1+\frac{\mu}{2}\right)}{\left(1+\eta_{A}\frac{\mu}{2}\right) \left(1+\eta_{B_i}\frac{\mu}{2}\right) \left(1+\eta_{A}\frac{\mu}{2}+\eta_{B_i}\frac{\mu}{2}-\eta_{A}\eta_{B_i}\frac{\mu}{2}\right)} ,
\end{equation}
where $Y_{0A(B_i)}$ is the background count rates, which quantifies the contribution of vacuum states, i.e., the dark count rate $p_d$ in our simulation. 
$e_0=1/2$ is the error rate of random noises, $e_d^{X}$ ($e_d^{Z}$) is the misalignment error rate of the $X$-basis ($Z$-basis), $\mu$ is the average number of photon pairs generated per pulse and $\eta_{A(B_i)} = \eta_{dA(B_i)} \cdot 10^{-\alpha L_{A(B_i)}/10}$, where $L_{A(B_i)}$ is the transmission distance between the network provider and the dealer $A$ (the player $B_i$).

By utilizing the post-matching technique~\cite{Lu2021EfficientQDS}, the gain of our protocol is the probability that both the dealer $A$ and each player $B_i$ receive and measure photons in the $X$-basis, which is $Q^X = p_x^2 Q_\mu$,
where $p_x$ is the probability that measuring photons in the $X$-basis. 
In the asymptotic limit of biased scheme~\cite{Lo2005EfficientQKD}, we have $Q^X \approx Q_\mu$. 
For the situation of perfect entangled photon pair source, which distributes one pure quantum state at a time, the gain and bipartite error rate are given as
\begin{equation}
	Q^X = Y_1 = \left[1-(1-Y_{0A})(1-\eta_{A})\right]\left[1-(1-Y_{0B_i})(1-\eta_{B_i})\right],
\end{equation}
\begin{equation}
	E_\mu^{X(Z)} = e_0 - \frac{\left(e_0-e_d^{X(Z)}\right)\eta_{A}\eta_{B_i}}{Y_1},
\end{equation}
where $Y_1$ is the yield for one photon pair, i.e., the probability of a coincidence detection event when the source emits one photon pair~\cite{Ma2007Entanglement}. 
Additionally, the gain for the QSS protocol based on perfect GHZ state source is 
\begin{equation}
	Q^X_{\mathrm{GHZ}} = \left[1-(1-Y_{0A})(1-\eta_{A})\right] \cdot \prod_{i=1}^{n-1} \left[1-(1-Y_{0B_i})(1-\eta_{B_i})\right].
\end{equation}
\end{widetext}

To analyze the marginal error rate in the $Z$-basis and the total error rate in the $X$-basis, we utilize the case of $n=3$ as an example and generalize the conclusions to the case of multiple parties. 
Here we list the measurement results and corresponding probabilities when the dealer $A$ obtains both logic 0 from the Bell states shared with the players $B_1$ and $B_2$ in the $Z$- and $X$-bases. 
For the other cases, where the dealer $A$ obtains logic $(0,1),(1,0),(1,1)$ in two bases, we should derive the same results. 
In order to align with the real situation, we assume that the classical operations performed with classical computer do not introduce any errors. 

\begin{figure*}[htbp]
\centering
\includegraphics[width=.8\linewidth]{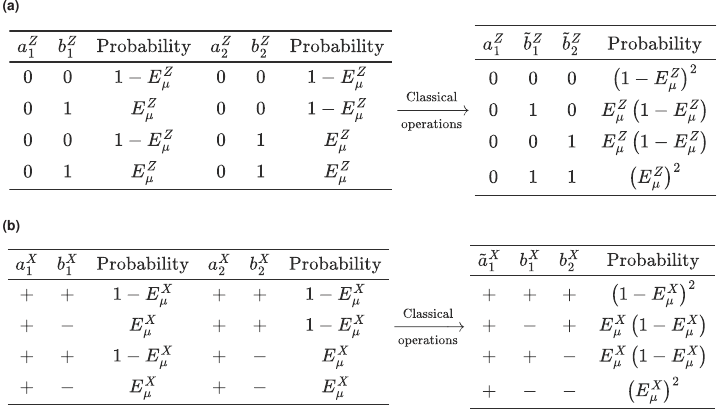}
\caption{The analysis of error rates in both $Z$- and $X$-bases, where the probability corresponds the possible measurement result. The error rates are the sum of corresponding probabilities. (a) For the $Z$-basis, the marginal error rate is the probability of the data inconsistency between the dealer $A$ and a single player $B_i$. (b) For the $X$-basis, the error rate is the probability that the data does not satisfy the GHZ-state correlation.}
\label{error rate}
\end{figure*}

For the $Z$-basis, only the discordant classical bits between the dealer $A$ and a single player $B_i$ contribute to the marginal error rate $E_{AB_i}^Z$. 
Based on the tables in Fig.~\ref{error rate}(a), the data in the second row and the fourth row contributes to the marginal error rate between $A$ and $B_1$, which is $E^Z_{AB_1} = E^Z_{\mu}\left(1-E^Z_{\mu}\right) + \left(E^Z_{\mu}\right)^2 = E^Z_{\mu}$.
Similarly, the marginal error rate between $A$ and $B_2$ is also $E^Z_{AB_2} = E^Z_{\mu}$, which is contributed by the data in the third row and the fourth row, so the marginal error rate $E_{AB_i}^Z$ of three parties is $E^Z_{\mu}$. 
We could consider the conclusion above in another way: due to the classical XOR operations do not introduce any errors, the marginal error rate of generated GHZ state between the dealer $A$ and the player $B_i$ can only be introduced by the error rate from bipartite entanglement distribution. 
Therefore, it is reasonable to generalize the conclusion to the case of $n$ parties, i.e., $E_{AB_i}^Z=E^Z_{\mu}$ holds for different numbers of participants.  

For the $X$-basis, owing to the GHZ-state correlation among classical data, only an odd number of error between the dealer $A$ and the player $B_i$ contribute to the total error rate $E^X$. 
For example, the tables in Fig.~\ref{error rate}(b) show that the data in the first row and the fourth row, which denotes no error and both the player $B_1$ and $B_2$ have errors with the dealer $A$, satisfy the GHZ-state correlation, i.e., $\tilde{a}_1^X=b_1^X \oplus b_2^X$. 
However, the data in the second row and the third row denotes only one player ($B_1$ or $B_2$) has error with the dealer $A$ and does not satisfy the GHZ-state correlation, so the total error rate is $E^X = 2E^X_{\mu}\left(1-E^X_{\mu}\right)$. 
We let $C_n^k = \frac{n!}{k!(n-k)!}$ be the binomial coefficient. 
The total error rate in the case of $n$-party should follow the same conclusion, therefore, we give the following total error rate 
\begin{equation}
	E^X = \sum_{j=0}^{\lfloor n/2 \rfloor -1} C_{n-1}^{2j+1} \left(E^X_{\mu}\right)^{2j+1} \left(1 - E^X_{\mu}\right)^{n-2j-2}	,
\label{eq:X basis error rate_1}
\end{equation}
where $\lfloor n/2 \rfloor$ denotes the floor function of $n/2$, $C_{n-1}^{2j+1}$ indicates that among all $n-1$ players, $2j+1$ players obtain discordant measurement results with the dealer $A$. 
Furthermore, the total error rate formula can be simplified as 
\begin{equation}
	E^X = \frac{1}{2} \left[1 - \left(1 - 2 E^X_{\mu}\right)^{n-1}\right], 
\label{eq:X basis error rate_2}
\end{equation}
and the proof is shown in below.
\begin{widetext}
\begin{equation}
\begin{split}
	E^X &= \sum_{j=0}^{\lfloor n/2 \rfloor -1} C_{n-1}^{2j+1} \left(E^X_{\mu}\right)^{2j+1} \left(1 - E^X_{\mu}\right)^{n-2j-2} \\
	&= \sum_{j=0}^{\lfloor n/2 \rfloor -1} C_{n-1}^{2j+1} \left(E^X_{\mu}\right)^{2j+1} \sum_{l=0}^{n-2j-2} C_{n-2j-2}^{l} \left(-E^X_{\mu}\right)^{l} \\
	&= \sum_{j=0}^{\lfloor n/2 \rfloor -1} \sum_{l=0}^{n-2j-2} C_{n-1}^{2j+1} C_{n-2j-2}^{l} \left(-1\right)^{l} \left(E^X_{\mu}\right)^{l+2j+1} \\
	&= \sum_{j=0}^{\lfloor n/2 \rfloor -1} \sum_{k=2j+1}^{n-1} C_{n-1}^{2j+1} C_{(n-1)-(2j+1)}^{k-(2j+1)} \left(-1\right)^{k-1} \left(E^X_{\mu}\right)^{k}, \quad (k=l+2j+1) \\
	&= \sum_{j=0}^{\lfloor n/2 \rfloor -1} \sum_{k=2j+1}^{n-1} C_{n-1}^{k} C_{k}^{2j+1} \left(-1\right)^{k-1} \left(E^X_{\mu}\right)^{k}, \quad \left(C_m^r C_{m-r}^{n-r} = C_m^n C_n^r\right) \\
\end{split}
\end{equation}
Here we change the order of summation and the range of the summation index, and utilize the properties of binomial coefficients $C_k^1 + C_k^3 + C_k^5 + \cdots = 2^{k-1}$ to simplify the formula. Moreover, $\lceil x \rceil$ denotes the ceiling function of $x$.
\begin{equation}
\begin{split}
	E_X &= \sum_{j=0}^{\lfloor n/2 \rfloor -1} \sum_{k=2j+1}^{n-1} C_{n-1}^{k} C_{k}^{2j+1} \left(-1\right)^{k-1} \left(E^X_{\mu}\right)^{k} \\
	&= \sum_{k=1}^{n-1} C_{n-1}^{k} \left( \sum_{j=0}^{\lceil k/2 \rceil -1} C_{k}^{2j+1} \right) \left(-1\right)^{k-1} \left(E^X_{\mu}\right)^{k} \\
	&= \sum_{k=1}^{n-1} C_{n-1}^{k} 2^{k-1} \left(-1\right)^{k-1} \left(E^X_{\mu}\right)^{k} \\
	&= -\frac{1}{2} \left[-1 + \sum_{k=0}^{n-1} C_{n-1}^{k} \left(-2E^X_{\mu}\right)^{k} \right] \\
	&= \frac{1}{2} \left[1 - \left(1 - 2 E^X_{\mu}\right)^{n-1}\right]
\end{split}
\end{equation}
\end{widetext}

\subsection{Simulation results}

\begin{figure}[h!]
\centering
\includegraphics[width=\linewidth]{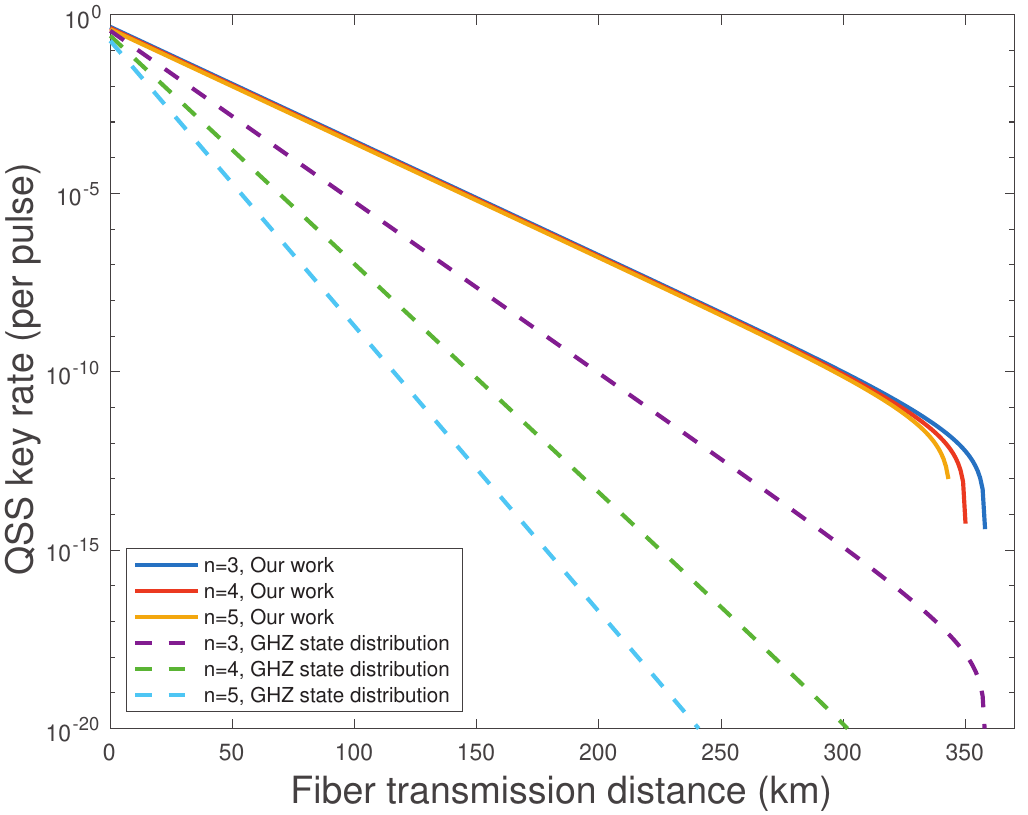}
\caption{
The simulation resuls of the comparison between our protocol and the QSS protocol distributing GHZ states directly under the perfect source. 
}
\label{Perfect source}
\end{figure}
\begin{figure*}[h!]
\centering
\includegraphics[width=\linewidth]{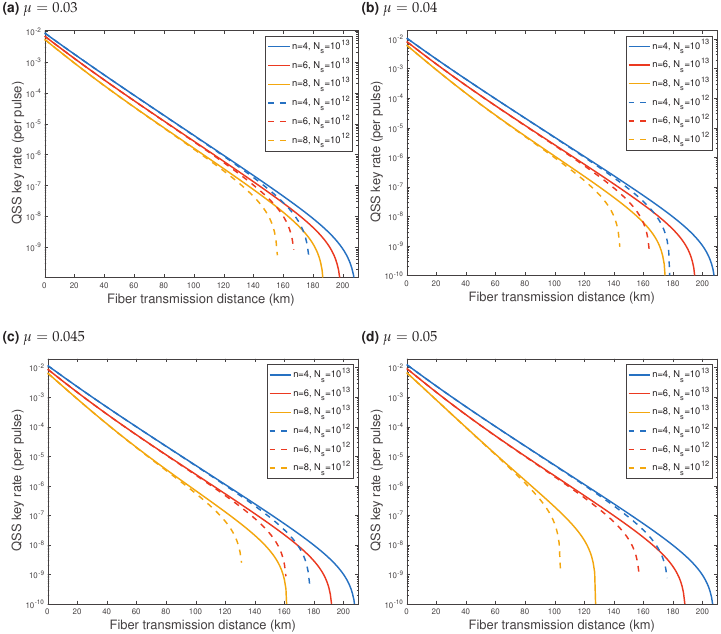}
\caption{The simulation results of key rates with the different average numbers of photon pairs generated by entangled photon pair source per pulse ($\mu=0.03,0.04,0.045,0.05$). The key rates are plotted under the different numbers of participants ($n = 4, 6, 8$) and the different numbers of signal distributed to each player ($N_s$ =  $10^{12}$, $10^{13}$). }
\label{Imperfect source}
\end{figure*}

\begin{table}[h!]
\centering
\caption{Simulation parameters. 
$e_d$ denotes the detector error. 
$\eta_d$ and $p_d$ are the detection efficiency and dark count rate. 
$\alpha$ is the attenuation coefficient with the $l$ kilometers fiber attenuation $10^{-\alpha l/10}$. 
$f_e$ is the error correction inefficiency. 
$p_x$ is the probability of $X$-basis measurement. 
$\epsilon_{\mathrm{c}}$, $\epsilon$ and $\epsilon^\prime$ are the failure parameters}
\begin{tabular}{ccccccccc}
\hline
$e_d$ & $\eta_d$ & $p_d$ & $\alpha$ & $f_e$ & $p_x$ & $\epsilon_{\mathrm{c}}$ & 3$\epsilon$ & 3$\epsilon^\prime$ \\
\hline
0.01 & 78\% & $10^{-7}$ & 0.16 & 1.12 & 0.9 & $10^{-10}$ & $10^{-10}$ & $10^{-10}$ \\
\hline
\end{tabular}
\label{Simulation_parameters}
\end{table}

Here, as a supplement for the main text, we utilize another group of parameters which are listed in Table~\ref{Simulation_parameters} to simulate the key rate of our protocol. 
Assume that the devices of all participants are the same and the transmission distances from the network provider to all participants are equal. 
We compare the performance between our protocol and the QSS protocol which directly distributes GHZ states under the perfect source. 
According to the simulation result in Fig.~\ref{Perfect source}, the key rate of our protocol is almost independent of the number of participants and offers a large advantage over the QSS protocol with GHZ states distribution, which is consistent with the result in the main text. 
With the help of the Bell state distribution and the post-matching technique~\cite{Lu2021EfficientQDS}, the gain of the QSS protocol is improved from $Q^X_{\mathrm{GHZ}} \propto \mathcal{O}\left(\eta^{n} \right)$ to $Q^X \propto \mathcal{O}\left(\eta^{2} \right)$, where $\eta = \eta_A = \eta_{B_i}$ ($i=1,2,\cdots,n-1$), thus the gain $Q^X$ of our protocol is independent of the number of participants $n$. 
Moreover, to address the internal participant attacks~\cite{Karlsson1999QSSParticipantAttack,Qin2007HBBCryptanalysis}, the marginal error rate in the $Z$-basis is taken into consideration, which is independent of $n$, however, the total error rate in the $X$-basis (Eq.~\ref{eq:X basis error rate_1} or Eq.~\ref{eq:X basis error rate_2}) is related with $n$. 
Therefore, the key rate of our protocol is almost independent of the number of participants. 

In addition, under the imperfect source, we simulate the key rates in finite-size regime with the different values of $\mu$, i.e., the average number of photon pairs generated by entangled photon pair source per pulse. 
Figure~\ref{Imperfect source} shows the finite key rate of our protocol with $\mu=0.03,0.04,0.045,0.05$. 
The key rate is almost independent of the number of participants with $\mu=0.03$, however, with the increase of the value of $\mu$, the key rate decreases significantly as the number of participants $n=8$. 
The main reason is that the multi-photon components in the signal increase with the increase of the average number of photon pairs generated per pulse, leading to a significant rise in the error rate, which results in the key rate decreasing markedly with the increase of the number of participants.

\bibliography{Entangle_QSS}

\end{document}